\documentclass[amssymb,aps,floats,twocolumn]{revtex4}
\usepackage{psfig}
\usepackage{color,graphicx,pstricks}
\draft
\begin{document}
\title{Domain Formation and Orbital Ordering Transition in a Doped Jahn-Teller Insulator}

\author{Sanjeev Kumar$^{1}$, Arno P. Kampf$^{1}$, and Pinaki Majumdar$^{2,3}$ }

\affiliation{$^{1}$~Institute of Physics, Theoretical Physics III, Center for Electronic
Correlations and Magnetism,\\
University of Augsburg, D-86135 Augsburg, Germany\\
$^2$~Institut Laue-Langevin, BP 156, 38042 Grenoble cedex 9, France\\
$^3$~Harish-Chandra  Research Institute,
Chhatnag Road, Jhusi, Allahabad 211 019, India
}

\begin{abstract}
The ground state 
of a double-exchange model for orbitally degenerate $e_g$ electrons 
with Jahn-Teller lattice coupling and weak disorder 
is found to be spatially inhomogeneous near half filling.
Using a real space Monte-Carlo method we show 
that doping the half-filled orbitally ordered insulator 
leads to the appearance of hole-rich
disordered regions in an orbitally ordered environment. The doping driven 
orbital order to disorder transition is accompanied by the emergence of
metallic behavior. We present results on transport and optical 
properties along with spatial patterns for lattice distortions and
charge densities, providing a basis for an
overall understanding of the low doping phase diagram of 
La$_{1-x}$Ca$_x$MnO$_3$. 
\end{abstract}
\vskip0.15cm

\maketitle

%--------------------------------------------------------------------------------
Hole-doped perovskite manganites, for example La$_{1-x}$Ca$_{x}$MnO$_3$ 
(LCMO), are well known for their colossal magnetoresistance (CMR) 
effect \cite{dagotto_book, tapan_book}. 
The `optimally doped' CMR compounds with $x \sim 0.3$
have been 
%in 
the focus of numerous theoretical studies and 
are qualitatively understood in terms of the interplay of the double 
exchange mechanism, electron-phonon interactions, and disorder 
in a single electronic band \cite{roder,brey,sk-pm-dhde,dag-prep}.
Less attention has been given to the low-doping regime, 
where the two-band character of 
manganites is crucial and in addition to  charge, spin, and 
lattice variables, the orbital degrees of freedom and their  
ordering become important. 
The undoped compounds are orbitally ordered (OO), A-type 
antiferromagnetic insulators \cite{A-type} with
large Jahn-Teller (JT) distortions of the MnO$_6$ octahedra \cite{murakami98},
which lift the degeneracy of the two Mn-$e_g$ levels. Electronic 
and cooperative lattice effects lead to a 
staggered ordering of the lattice distortions and the orbital occupancies.
Upon doping, the antiferromagnetic insulator evolves 
into a ferromagnetic insulator, 
with  weakened orbital order, and eventually undergoes a transition to an orbitally 
disordered  ferromagnetic metal (OD-FM-M) \cite{biotteau01}.
Despite the achieved progress towards an understanding of magnetic and 
orbital ordering in the 
undoped compounds \cite {dagotto99}, efforts for analyzing the doping driven 
transition from the orbitally ordered insulating to the orbitally disordered metallic 
phase have remained limited. 
A simple view of this transition rests on an entirely classical picture
in terms of random fields introduced by the doped holes \cite{millis96}.

The critical hole doping $x_{OD}$ for the loss of orbital order
is close to the doping $x_{IMT}$ for the 
insulator-metal transition (IMT) 
in LCMO, 
where  $x_{IMT} \sim 0.22$ \cite{vanaken03}. 
In lower bandwidth 
materials like Pr$_{1-x}$Ca$_x$MnO$_3$ (PCMO)
the insulating phase persists to even larger hole concentrations \cite{pcmo}.
NMR and neutron scattering experiments suggest,  
that the doping regime $x \lesssim x_{IMT}$ is spatially
inhomogeneous in LCMO 
with coexisting `hole poor' orbitally ordered and `hole rich'
orbitally disordered regions \cite{papa03-hennion98}, and 
the observation of
confined  `spin waves' confirms the existence of 
magnetic clusters on the nanoscale \cite{hennion05}. 
It has remained unclear  
how the JT insulator evolves from the   
homogeneous OO state  at $x=0$ to the 
homogeneous OD metal  at
optimal doping through an intermediate inhomogeneous state.

In this Letter we present results on a two-band double-exchange 
model with electron-lattice
coupling and disorder 
in two dimensions (2D) using a real-space technique. 
We provide a description for 
the doping driven loss of orbital order
and  the detailed doping vs. temperature phase diagram at
intermediate electron-lattice coupling,  appropriate to LCMO.
Results for charge transport and spectral properties are presented, which characterize
the metal-insulator transitions. Real-space structures for the 
inhomogeneous state of the hole doped insulator allow us to follow the emergence
of orbitally disordered domain walls
at low doping and their eventual percolation.

Specifically, we consider a two-band model for itinerant $e_g$ 
electrons coupled to JT lattice distortions
and to localized $S=3/2$ $t_{2g}$ spins in the
presence of substitutional disorder, described by 
the Hamiltonian:
\begin{eqnarray}
H &=& \sum_{\langle ij \rangle \sigma}^{\alpha \beta}
t_{ij}^{\alpha \beta} 
c^{\dagger}_{i \alpha \sigma} c^{~}_{j \beta \sigma} 
+ \sum_i (\epsilon_i -\mu)n_i
 - J_H\sum_i {\bf S}_i {\bf \cdot} { {\mbox {\boldmath $ \sigma $}} }_i 
\cr
&& + J_S\sum_{\langle ij \rangle} {\bf S}_i \cdot {\bf S}_j 
+ \lambda \sum_i {\bf Q}_i {\bf \cdot} {   {\mbox {\boldmath $ \tau $}} }_i 
+ {K \over 2} \sum_i |{\bf Q}_i|^2 .~ ~ ~ ~ ~ ~ ~
\end{eqnarray}

\noindent
The magnetic properties arise from the competition between
the Hund's rule coupling $J_H$ 
driven double exchange and the antiferromagnetic
superexchange $J_S$ between the $t_{2g}$ core spins ${\bf S}_i$.

In Eq. (1), $ c^{}$ and $c^{\dagger}$ are annihilation and creation operators for
$e_g$ electrons and
$\alpha$, $\beta $ are summed over the two Mn-$e_g$ orbitals
$d_{x^2-y^2}$ and $d_{3z^2-r^2}$, which are labelled $(a)$ and $(b)$ in what follows.
$t_{ij}^{\alpha \beta}$ are the hopping matrix elements between
$e_g$ orbitals on nearest-neighbor sites and have the cubic perovskite specific form:
$t_x^{a a}= t_y^{a a} \equiv t$, 
$t_x^{b b}= t_y^{b b} \equiv t/3 $,
$t_x^{a b}= t_x^{b a} \equiv -t/\sqrt{3} $,
$t_y^{a b}= t_y^{b a} \equiv t/\sqrt{3} $ \cite{dagotto_book}, where
$x$ and $y$ denote the spatial directions on a square lattice.
The disorder is modelled by random on-site potentials $\epsilon_i$,
with equally probable values $\pm \Delta$.
The $e_g$-electron spin is ${\sigma}^{\mu}_i= 
\sum_{\sigma \sigma'}^{\alpha} c^{\dagger}_{i\alpha \sigma} 
\Gamma^{\mu}_{\sigma \sigma'}
c_{i \alpha \sigma'}$, 
where $\Gamma^{\mu}$ are the Pauli matrices.
$\lambda$ denotes the strength of the JT coupling between the distortion 
${\bf Q}_i = (Q_{ix}, Q_{iz})$ and  
the orbital pseudospin
${\tau}^{\mu}_i = \sum^{\alpha \beta}_{\sigma}
c^{\dagger}_{i\alpha \sigma} 
\Gamma^{\mu}_{\alpha \beta} c_{i\beta \sigma}$ \cite{dagotto_book}. 
$K$ controls the lattice stiffness, and $\mu$ is the chemical potential.

We set $t=1$ as the reference energy scale. 
In the manganites $J_H \gg 1$, and we adopt the frequently used limit
$J_H \rightarrow \infty$, which retains the 
essential physics \cite {dagotto_book}. We use $J_s = 0.05$ throughout, which
is estimated from the N\'eel temperature for CaMnO$_3$,
where antiferromagnetism is purely superexchange driven.
The parameters $\lambda$ and  $\Delta$
will be selectively explored. 
The spins are assumed to be classical
unit vectors, $\vert {\bf S}_i \vert =1$; quantum effects in the lattice variables are
not considered,
and the stiffness is set to $K=1$.
In the limit $J_H \rightarrow \infty$ the 
spin of the $e_g$ electrons is tied to the orientation of the local 
core spin leading to a two-orbital `spinless' fermion model
with core spin configuration dependent hopping amplitudes \cite{dagotto_book}.

Replacing a fraction $x$ of rare earth ions with
$2^+$ cations in the parent manganites affects the mean 
A-site ionic radius $r_A(x)$ as well as its variance $\sigma_A(x)$.
The varying $r_A$ modifies the electronic hopping amplitude,
and hence the $\lambda/t$ ratio, while $\sigma_A$ controls the disorder strength $\Delta$.
In most of what follows we set $\lambda=1.6$, which reproduces
the transport gap $\sim 0.4$eV (if we assume  $t \sim 0.2$eV)
in LaMnO$_3$ estimated from the activated resistivity behavior \cite{palstra}.
We set $\Delta=0.4$ as a typical value for weak disorder, and explore the
doping and temperature dependence. 
Naturally the amount of disorder
depends on the doping level,
but this variation is not addressed here.
  
The model defined in Eq. (1) has been studied earlier
using mean-field methods, as well as exact diagonalization (ED) based Monte-Carlo (MC)
simulations \cite{kilian-dag98}. While the mean-field approximation excludes by construction
the possible existence of
inhomogeneous phases, the accessible system sizes within ED-MC are too small ($\sim$ 100 sites)
to explore spatial clustering effects, the orbital order to disorder transition, or the IMT itself.
Here we use the travelling cluster approximation (TCA), 
which readily allows access to systems of $\sim 1000$ sites
to anneal the classical spin and lattice variables.
ED of the full fermionic Hamiltonian is used 
only for computing the electronic 
quantities in the TCA-generated classical
configurations. This method has been benchmarked before \cite{TCA} 
and applied to a one-band version of Eq. (1) \cite {sk-pm-dhde}.

Fig. 1 summarizes our results for the OO-OD transition, the
core-spin magnetism, and the doping and temperature driven IMT. 
Panel (a) shows the $x-T$ phase diagram
for $\lambda=1.6$ and $\Delta=0.4$.
Squares mark the ferromagnetic (FM) to paramagnetic (PM) 
crossover and
circles denote the orbital ordering transition temperatures
$T_{OO}$ as inferred from the temperature
dependence of the ${\bf q} = (\pi, \pi) \equiv {\bf q}_0 $ component of the lattice structure factor,
$D_{Q} ({\bf q})$~ $ = N^{-2} \sum_{ij} \langle
{\bf Q}_i \cdot {\bf Q }_j \rangle_{av} ~ ~ e^{-i{\bf q}\cdot ({\bf r}_i - {\bf r}_j)} $
shown in panel (c). Here and below $\langle ... \rangle_{av}$ 
denotes the combined average over thermal equilibrium configurations and 
over the realizations of quenched disorder.
The I-M boundary is obtained
from the sign of the slope of the resistivity $\rho(T)$ (see Fig. 2(a)).
At $T=0$  the system is ferromagnetic 
at all doping levels, despite the presence of the antiferromagnetic superexchange coupling $J_S$;
the doping driven OO-OD transition at $x_{OD} \sim 0.22$ occurs
close below the IMT. For $x < x_{OD}$ the system loses either ferromagnetic
order (for $x \rightarrow 0$) or orbital order first
(for $x $ near $ x_{OD}$) with increasing $T$, and for
$T \gtrsim 0.1$ all long range order is lost
leading to an orbitally disordered paramagnetic insulator (OD-PM-I).

% ------------------------------------------------------------------------------
\begin{figure}[t!]
\centerline{\psfig{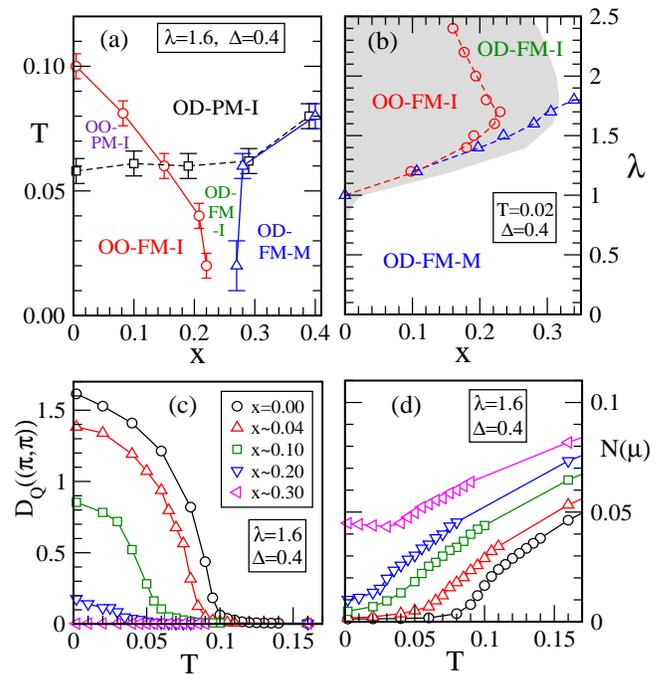}}
\caption{(Color online)~ (a)~ $x-T$ phase diagram at $\lambda=1.6$.
(b)~ $x-\lambda$ phase diagram at $T=0.02$. The disorder strength
is $\Delta=0.4$. The shaded region in (b) indicates phase separation. 
OO (OD) refers to an orbitally ordered (disordered) phase; FM (PM) denotes a
ferromagnetic (paramagnetic) state and M (I) indicates metallic (insulating) character.
$T$ dependence of (c) the lattice structure factor $D_Q((\pi, \pi))$
 and (d) the density of states at the chemical potential
$N(\mu )$ for different $x$. }
\end{figure}
% ------------------------------------------------------------------------------

A broader perspective for the OO-OD transition
is obtained from the low temperature $\lambda-x$ 
phase diagram for $\Delta=0.4$, 
shown in Fig. 1(b). 
For our choice of $J_S$ the system is FM  over the entire 
selected parameter range.
In the clean limit, $\Delta=0$, the major feature is 
phase separation (PS), as indicated by the grey shaded area,
between the OO-FM-I at $x=0$ and the OD-FM-M for $x > 0$.
PS is identified from the existence of a jump in the average hole density
upon varying the chemical potential. In the presence of
disorder the PS range is replaced by inhomogeneous phases.
We identify three distinctly different regimes with respect to $\lambda$:
$(i)$~For $\lambda \lesssim  1$ orbital ordering is absent
and the system is metallic over the entire doping range.
$(ii)$~When $1 < \lambda < 1.6 $ the system sustains orbital order even away from
$x=0$, and the orbital order and the insulating character is lost at a critical
doping $x_{OD}(\lambda) \approx x_{IMT}(\lambda)$, which reflects
that the insulating character is intimately related
to the staggered ordering of the lattice distortions.
$(iii)$~For $ \lambda > 1.6 $ the system first loses 
orbital order and subsequently becomes metallic at higher doping, so  $x_{IMT} > x_{OD}$.
In fact, for $\lambda  \gtrsim 2$ the system does not become metallic
even for $x > 0.3$. This is the strong coupling regime where single
electrons can be `self-trapped' and orbital ordering is
not a prerequisite for the insulating behavior.
If we assume that $\lambda/t \sim 1.6$ is 
appropriate to LCMO, the
phase diagram in Fig. 1(a) agrees remarkably well with the experimental results reported in Ref. \cite {vanaken03}.
The lower bandwidth materials would correspond to larger $\lambda/t$,
and indeed we find that the insulating character persists to much larger $x$ for larger $\lambda$,
consistent with the experiments on PCMO \cite{pcmo}. We remark that some aspects of the 3D 
manganite physics are
not contained in our 2D calculation, {\it e.g.} we can not address the crossover from an
A-type antiferromagnet to a FM state; the present results should thus be compared to 
the physics of a single plane of the real materials only.
 
Fig. 1(c) shows the temperature dependence of   
$D_{Q} ({\bf q}_0)$,
which is a measure of the staggered
ordering tendency of the JT distortions,
for different hole concentrations.
Since the lattice distortions are coupled to the orbital 
pseudospin, the same 
tendency is 
transferred to the analogously defined orbital structure factor 
$D_{\tau}({\bf q}_0)$. Therefore $D_{Q}({\bf q}_0)$ serves as an 
indicator for the staggered ordering of both, the lattice 
distortions and the orbital pseudospin. Staggered order is
found only in the $Q_x$ component of the distortions.

In the undoped case the onset of orbital order 
is accompanied by the opening of a gap around the Fermi level
in the density of states (DOS) 
$ N(\omega) = N^{-1} \left \langle \sum_n 
\delta(\omega- \epsilon_n\{{\bf Q},{\bf S}\}) \right \rangle_{av} $, where
$\epsilon_n\{{\bf Q},{\bf S}\}$ denote the 
single particle eigenvalues 
for a configuration $\{{\bf Q},{\bf S}\}$
of the classical variables. 
Fig. 1(d)  tracks the temperature dependence
of $N(\mu )$ for the same choice of $x$ values as in Fig. 1(c).
$N(\mu )$ vanishes at low temperatures for $x \lesssim 0.10$, 
indicative for insulating behavior  
but retains a small finite value for $x \gtrsim  0.10$; the latter
is in fact related to 
a pseudogap structure in the DOS and suggests a possibly 
`metallic' state (see Fig. 2).

% ------------------------------------------------------------------------------
\begin{figure}[t!]
\centerline{\psfig{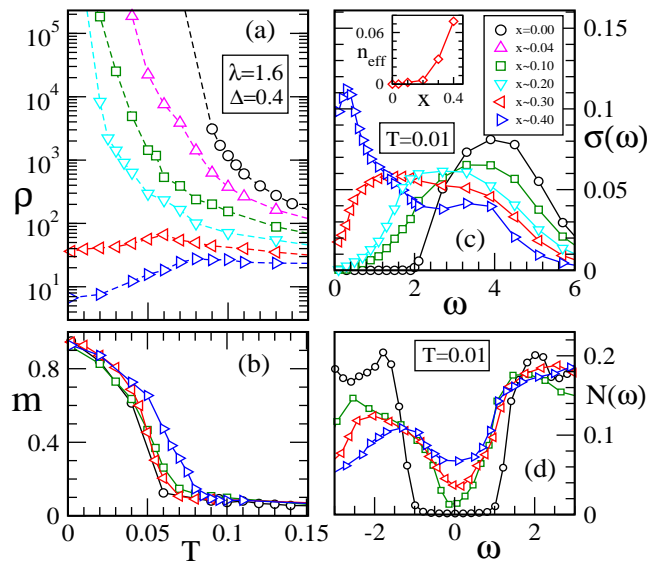}}
\caption{(Color online)~ (a) Resistivity $\rho $ on a logarithmic scale in
units of  $\hbar/\pi e^2$ 
and (b) magnetization $m$ as a function of temperature
for different $x$. The low temperature 
(c) optical conductivity $\sigma(\omega)$
and (d) density of states $N(\omega)$ for varying $x$. 
The results are for 
$\lambda = 1.6 $ and $\Delta = 0.4$. 
A Lorentzian broadening of 0.02 is employed for the DOS calculations.}
\end{figure}
% ------------------------------------------------------------------------------

In order to discuss the implications for transport 
we compute the optical conductivity $\sigma(\omega)$ using 
the Kubo formula with the exact eigenstates 
\cite{transport}. The resistivity $\rho$ is approximated
by the inverse of $\sigma(\omega_{min})$, where $\omega_{min} = 20t/N
\sim 0.03t$
is the lowest reliable frequency 
scale for $\sigma(\omega)$ calculations on our $N=24^2$ system.
Fig. 2(a) shows the temperature dependence of $\rho$ for different $x$.
For $x \leq 0.2$ there is a sharp rise in $\rho(T)$ at the onset temperature for 
orbital ordering $T_{OO}$.
For $x \sim 0.3$ and $x \sim 0.4$, however,  there is a downturn 
in $\rho(T)$ upon cooling.
We correlate this behavior with the 
temperature dependence of the magnetization $m(T)$  
defined via  
$ m^2 = \left \langle (N^{-1} \sum_i {\bf S}_i )^2 
\right \rangle_{av} $, shown in Fig. 2(b). 
For all values of $x$ a para- to ferromagnet transition is observed upon cooling,
with the Curie temperature $T_c$ inferred from the 
inflection point in $m(T)$. Note that $T_c$ does not increase with
increasing $x$, contrary to the simple double exchange scenario \cite{calderon}.
Fig. 2(c) shows the optical conductivity at low temperature 
for different $x$.
For $x \lesssim 0.2$ the low-frequency spectral weight is 
strongly suppressed, which is evident in the effective carrier density $n_{eff} (\omega,x)
= \int_0^{\omega} \sigma(\omega') d \omega'$ at $\omega
=1 $, shown in the inset.
For $x=0.3$, 
$\sigma(\omega)$ 
remains finite at the lowest attainable $\omega$,
but the response is non-Drude like, while at
$x=0.4$ a more conventional optical
response emerges. Nevertheless, even at $x=0.4$, $n_{eff}$ is strongly
suppressed compared to the naive electron count.
This is consistent with our observation in the disordered one-band model \cite{sk-pm-dhde}
and appears as a generic feature of the interplay of disorder and electron-lattice coupling.
Fig. 2(d) shows how the DOS evolves from the 
`clean gap' at $x=0$ through the low $x$ pseudogap to the 
high $T_c$ `metallic' regime. The clean gap at $x=0$, which 
originates from the nesting instabilities at weak coupling \cite{khomskii} and
effective repulsions between self-trapped electrons at strong coupling,
is stable in the presence of weak disorder \cite{sk-apk-pm}.
From the combination of $N(\mu )$, $\rho(T)$, and $\sigma(\omega)$ 
we conclude that the electron system at $T=0$ has
an 'IMT' near $x=0.25$.

For a more microscopic understanding of these results, we have examined 
the doping evolution of the real space patterns for the local charge density $n_i$ and
the local spatial correlations of the lattice variables 
$C_Q^i = \frac{1}{4}\sum_{\delta} 
{\bf Q}_i \cdot {\bf Q}_{i+\delta} $ for a representative disorder realization,
where $\delta$ is summed over
the nearest neighbors of site $i$.
The $n_i$ in the upper row in Fig. 3 show 
(with white marking the `hole poor' and
black the `hole rich' regions) that the
doped holes at low $x$ 
lead to a strong density variation, although the holes are
not `site localized'. The hole positions are spatially correlated
with a short range charge order pattern. The 
spatial pattern is filamentary, rather than `puddle-like',
with the linear structures connecting up for $x \sim x_{OD}$.
% ------------------------------------------------------------------------------
\begin{figure}[t!]
\centerline{\psfig{figure=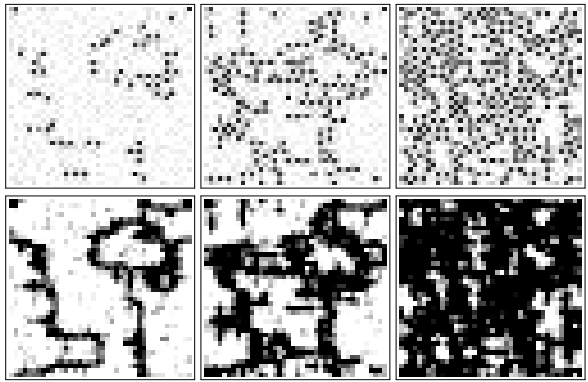,width=8.5cm,height=5.4cm,angle=0}}
\caption{Spatial patterns at $x=0.04, 0.11$ and $0.19$ (left, middle and
right). 
Top row: charge density $n_i$, grayscale covers the range from 0.2 (black) to 1 (white). 
Bottom row: local correlations of the lattice distortions $C_Q^i$ (see
text). Grayscale from -1.5 (white) to 0 (black). 
All results are for a specific disorder realization
on a $40 \times 40$ lattice,
$T=0.01, \lambda=1.6$ and $\Delta=0.4$}.
\end{figure}
% ------------------------------------------------------------------------------
The $C_Q^i$ pattern is understood from the hole locations, 
and the above discussed supression of $D_Q({\bf q}_0)$ arises from the loss
of OO in the vicinity of the holes as well as
from the presence of antiphase domains separated by the `hole rich' domain walls. 

Although local Coulomb interactions were not explicitly included in our model 
analysis their effects are nevertheless partially captured.
{\it E.g.} the
large $J_H$ avoids double occupancy of a single orbital and therefore acts like an 
{\it intra-orbital} Hubbard repulsion.
The JT polaron binding energy, $\sim \lambda^2/2K$, has effects
similar to an {\it inter-orbital} Hubbard repulsion since it
prefers one $e_g$ electron per site \cite{dagotto99}. Therefore we do not expect
qualitative changes in our phase diagrams if explicit electron-electron 
interactions were included. The $T_{OO}$ scale however will be affected. 
The cooperative nature of the lattice distortion will also enhance
$T_{OO}$ as we have checked. 
The critical doping for the OO-OD transition, however, does not seem
to be significantly affected by cooperative effects,
and neither are the spatial patterns for $x \sim x_{OD}$ \cite {sk-apk-pm}.

Naturally the 'IMT' that we observe in our 2D model with disorder is to be understood as a crossover from an 
insulating phase to a weakly localized (WL) phase with a finite density of states at the Fermi level.
The localization effects become apparent by studying progressively larger lattices,
and we observe the expected slow growth of the computed resistivity.
In 3D the gapped insulator to WL crossover is expected to become a genuine IMT.

In summary, our results on the 2D Jahn-Teller double exchange
model reveal a doping driven transition from an orbitally 
ordered insulator to an 
orbitally disordered ferromagnetic metal in agreement with the experiments on 
La$_{1-x}$Ca$_x$MnO$_3$.
In the orbitally disordered regime, the
system undergoes a thermally-driven transition from a ferromagnetic metal to a 
paramagnetic insulator, characteristic of the CMR materials. The intermediate 
inhomogeneous phase, with coexisting orbitally 
ordered and orbitally disordered regions, 
allows a natural interpretation of the neutron scattering and NMR data in LCMO.

SK and APK gratefully acknowledge support by the 
Deutsche Forschungsgemeinschaft through SFB 484.
PM was supported by Trinity College and the EPSRC, UK (at Cambridge 
University), and the Royal Society, UK
(at Oxford University). Simulations were performed on the Beowulf 
Cluster at HRI.

{}

\end{document}